\documentclass[12pt]{iopart}
\usepackage{iopams}  
\usepackage{graphicx}% Include figure files
\usepackage{dcolumn}% Align table columns on decimal point
\usepackage{bm}% bold math
\def\nbOne{\ \!\hbox{{\rm 1$\hskip-2.7pt$l}}}
\def\nbM{\ \!\hbox{{\rm I$\hskip-2.2pt$M}}}

\newcommand{\bfa}{{\cal W}\!\!\!\!\!\!{\cal W}}

\begin{document}

\title[Perfect spin filtering device]{Perfect spin filtering device through a Mach Zehnder interferometer in GaAs/AlGaAs electron gas}

\author{Alexander L\'opez$^1$, Ernesto Medina$^{1,2,3}$, Nelson Bol\'ivar$^2$ and Bertrand Berche$^3$}

\address{$^1$Centro de F\'\i sica, Instituto Venezolano de Investigaciones Cient\'ificas, Apartado 21874, Caracas 1020-A, Venezuela}
\address{$^2$Departamento de F\'\i sica, Universidad Central de Venezuela, Caracas, Venezuela}
%\ead{custserv@iop.org}
\address{$^3$Statistical Physics Group, P2M, Institut Jean Lamour, Nancy Universit\'e, BP70239, F- 54506 Vand\oe uvre les Nancy, France}
\begin{abstract}
A Spin filtering device through quantum spin interference is addressed, in two dimensions, in a GaAs/AlGaAs electron gas that has both Rashba and Dresselhaus spin-orbit couplings and an applied external magnetic field. We propose an experimentally feasible electronic Mach Zehnder Interferometer and derive a map, in parameter space, that determines perfect spin filtering conditions. We find two broad spin filtering regimes, one where filtering is achieved in the original incoming quantization basis, that takes advantage of the purely non-Abelian nature of spin rotations, and the other, where one needs a tilted preferential axis to observe the polarized output spinor. Both solutions apply for arbitrary incoming electron polarization and energy, and are only limited in output amplitude by the randomness of the incoming spinor state. A full account of beam splitter and mirror effects on spin renders solutions only on the tilted basis, but encompasses a broad range of filtering conditions.

\end{abstract}

%Uncomment for Submitted to journal title message
\submitto{\JPCM}
\maketitle

\section{Introduction}
The Rashba and Dresselhaus SO interactions arise in materials which lack either structural or bulk inversion symmetry, respectively\cite{Rashba,Dresselhaus,winkler}. These two kinds of interactions have recently been given a great deal of attention due to their potential role in the generation and manipulation of spin polarized currents, spin filters\cite{Nitta,Ionicioiu,Hatano,SHChen}, spin accumulation\cite{SarmaReview}, and spin optics\cite{BalseiroUsaj}. 

A reformulation of the spin-orbit coupling Hamiltonian in terms of non-Abelian gauge fields\cite{ryder} was explicitly given in ref. \cite{Rebei,Jin,Leurs,Medina} where the SO interaction is presented as a  $SU(2)\times U(1)$ gauge theory. As the Yang-Mills gauge theory is well understood and is the underpinning of well established theory, enormous insight can be brought upon new problems. Such gauge point of view, in more general terms, has been known for some time\cite{Goldhaber,Mineev,Frohlich}. This formulation is very revealing, since the consistent gauge structure of the theory becomes obvious and the physics of spin currents, persistent currents and color diamagnetism\cite{Tokatly} can be understood in a manner analogous to the well known $U(1)$ gauge theories. A consistent  $SU(2)\times U(1)$ gauge approach was presented in reference \cite{Leurs,Medina} where it was found that for the Pauli type Hamiltonians (including Rashba and 2 dimensional reductions of the Dresselhaus Hamiltonian), Gauge Symmetry Breaking (GSB) is necessarily built into the theory and leads to vanishing of the spin conductivity in constant electric fields\cite{Medina}. In addition, the Yang Mills interpretation of the Rashba and Dresselhaus SO interactions renders the associated gauge fields real, with topological consequences analogous to the Aharonov Casher effect\cite{Leurs,Medina}. 

Recent proposals were recently reported for the construction of perfect spin filters based on active Rashba spin orbit  media\cite{Hatano}, ballistic spin interferometers\cite{Koga} and the analysis of the persistent spin helix\cite{Bernevig2,SHChen}, where the Yang Mills gauge point of view is advantageous. Here we readdress the problem of spin filtering by interferometry in a quasi two dimensional system, and make connection to an experimentally feasible test of these ideas through an electronic Mach Zehnder interferometer (MZI) within Rashba and Dresselhaus media. Recent proposals contemplating this setup as an spin intereference device include quantum logic gates\cite{ZulickeAlone}, bit controlled Stern-Gerlach devices\cite{Ionicioiu} and tunable entanglement\cite{SignalZulicke}. Our analysis, within this setup, enables us to obtain exact conditions for spin filtering which can be achieved by tuning appropriate experimental parameters. Such conditions for spin filtering greatly generalize previous special situations where the spin polarization is a conserved quantity\cite{Ting}, and show new possibilities for spin filtering beyond previous approximate treatments.

The structure of the paper is as follows. First we consider the Hamiltonian with both Rashba and Dresselhaus interactions for a two dimensional electron gas (2DEG) including a magnetic flux described by a $U(1)$ gauge field. Following the approach given in ref. \cite{SHChen}, we rewrite the Rashba and Dresselhaus contributions in terms of a Yang Mills gauge field and review how this approach leads to the introduction of a GSB term analogous to the Proca term for the Maxwell field. Then, we propose an interference setup in the form of an electronic MZI where the electron's spin transport is modulated due to the presence of Rashba and Dresselhaus active media. We derive the conditions for perfect spin filtering that are applicable independently of the incoming spin state and the full energy range of the injected electrons. Finally, we give some concluding remarks.

\section{Spin-Orbit scattering for two dimensional electron gas}
We consider a two dimensional system consisting of non interacting electrons subject to both Rashba and Dresselhaus spin orbit interactions. In addition, one can apply an external transverse magnetic flux $\Phi_B$ described by a $U(1)$ gauge vector potential $\bf A$.  Two recent works have shown how to measure and control the Rashba and Dresselhaus parameters using gate voltages in two dimensional GaAs/AlGaAs electron gas\cite{MillerGoldhaberGordon,Studer}. It is striking that one can achieve SO magnetic fields of 2-3 mT. The SO physics beautifully follows an extended weak localization theory that allows for a detailed access to the material parameters. 

One can address the two dimensional GaAs/AlGaAs electron gas by a single particle Hamiltonian including the previously described couplings by
\begin{equation}\label{Hamiltonian}
H= \frac{{\bf \Pi}^2}{2m^*} + V - \alpha (\Pi_x\sigma^y-\Pi_y \sigma^x)- \beta(\Pi_y\sigma^y-\Pi_x\sigma^x)+ \frac{\hbar \omega_B}{2}\sigma^z, \label{H1}
\end{equation}
where ${\bf \Pi}={\mathbf p}+e{\mathbf A}$, $-e$ and $m^*$ are the electron's charge and effective mass, $V$ a substrate lattice potential that can be assumed periodic, $\boldsymbol \sigma$ is a vector of Pauli matrices, and $\alpha$ and $\beta$ are material-dependent parameters characterizing the Rashba and Dresselhaus interactions, respectively. The last term is the Zeeman energy. The term linear in $k$ describing the Dresselhaus interaction results from averaging a cubic in $k$ contribution (for the bulk) in the confining direction and neglecting other cubic terms in the strong lateral confinement situation\cite{Halperin}. In the rest of this work we ignore the effect of the Zeeman term in the limit of small magnetic fields (a few flux quanta through a ${\rm 200}\times {\rm 200} \mu {\rm m}^2$ area) such that the spin orbit energy is much larger than the Zeeman energy\cite{Takayanagi}. According to measured parameters in ref. \cite{MillerGoldhaberGordon} the SO energy for an GaAs/AlGaAs electron gas is 5 orders of magnitude greater than the Zeeman energy for the proposed field strengths. This way the external magnetic field results in strong phase effects through the vector potential but no appreciable precession occurs due to the Zeeman term. Nevertheless, we will see that there are spin filtering scenarios for the device even for zero external magnetic field.

Following \cite{Hatano,SHChen}, we introduce a spin dependent (non-Abelian) gauge field $\bfa$ whose components are given by
\begin{equation}
\frac{g}{m^*}\bfa^a\tau^a 
= (\beta\tau^x-\alpha\tau^y)\hat {\mathbf x}+(\alpha\tau^x-\beta\tau^y)\hat {\mathbf y},
\end{equation}
with $\tau^a=\sigma^a/2$, and $g/\hbar$ is the $SU(2)$ coupling constant. Using this gauge field we can rewrite equation (\ref{H1}), having ignored the Zeeman contribution, in the form 
\begin{eqnarray}
H=\frac{\left ({\mathbf p}+e{\mathbf A}+g\bfa^a\tau^a\right)^2}{2m^*}+eA_0-\frac{g^2\bfa^a\cdot\bfa^a}{8 m^*}.
\label{Pauli2}
\end{eqnarray}
\begin{figure}
\begin{center}
\includegraphics[width=8 cm]{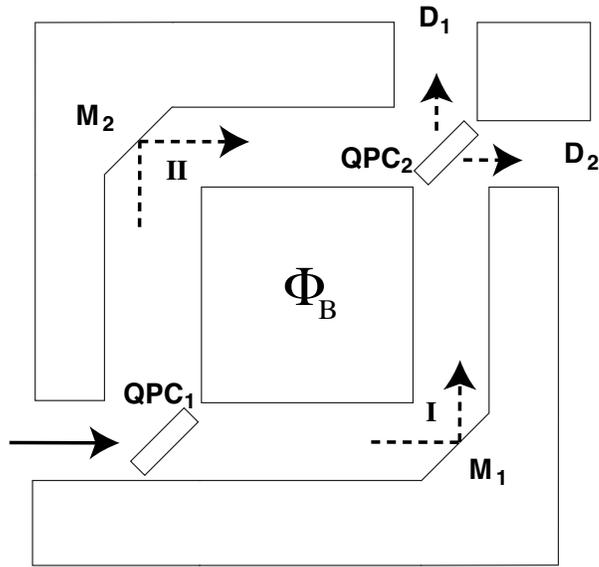}
\end{center}
\caption{Sketch of the electronic Mach Zenhder interferometer setup. The arms of the square are made of active SO Rashba and Dresselhaus media. The beam splitters are implemented through two Quantum Point Contacts (QPCs). There is a magnetic flux $\Phi_B$ through the square.}
\label{fig1}
\end{figure} 
The first term describes the total kinetic energy taking into account the contribution from the regular vector potential due to an external magnetic field and the non Abelian gauge field. The second term is the background lattice potential whereas the third term represents a gauge symmetry breaking contribution similar to the field originally discussed in references \cite{Leurs,Medina,Comment} responsible for rendering the spin currents physical.  

\section{Electronic Mach Zehnder spin interferometer}
A device configuration that allows us to address the problem of spin filtering in a gauge independent\cite{Medina} manner is the Mach Zehnder Interferometer (MZI). The setup for an MZI is sketched in (figure \ref{fig1}). Here we are interested in determining the resulting amplitude $\Psi_{D_i}$ at detector $D_i$, with $i=1,2$ and to find the conditions for perfect spin filtering \cite{Hatano} at either detector. There is an interesting issue that must be discussed regarding spin $1/2$ filtering. If the state at the input is a pure state spinor of spin $1/2$, the electron is polarized on some indeterminate axis, in principle random, coming from the Fermi sea of the input conductor. If one could find this axis for every electron extracted then one would have a perfect spin filter for each electron. Nevertheless the resulting current is unpolarized. We thus define the spin filter as one acting on any entering (pure state) polarization and returning a polarized state along a definite axis. This approach will serve to build a polarized spin current.

The relevant processes within the interferometer are described as follows (see figure \ref{fig1}): Single electrons are assumed to be extracted from the Fermi sea as pure states $\Psi_0={\psi^+_0\choose\psi^-_0}$. The electrons then pass through a beam splitter that can be implemented by a combination of Quantum Point Contacts\cite{OliverYamamoto} the first of which we label ${\rm QPC_1}$ described by a $4\times 4$ scattering matrix $S_1$ that mixes spin orientations on perpendicular reflection, while it is diagonal for direct (no change in direction) transmission\cite{Yamamoto}. Mixing of spin orientations occurs at all reflections (including mirrors) due to changes in direction of the electron ${\bf k}$ vector within spin-orbit active media that changes the orientation of the implied wavevector-dependent magnetic field. Furthermore, as we consider both Rashba and Dresselhaus interactions, we need to derive general reflection conditions at the beam splitters and mirrors. In reference \cite{Yamamoto}, this was done for Rashba assuming that small enough spin-orbit strength would yield only a small divergence of the reflected spin states in a ${\bf k}$ dependent basis. Surprisingly, when only the Rashba interaction is involved, the reflection matrix depends only on the incident angle and the reflection coefficient. On the other hand, if both Dresselhaus and Rashba are included, this is no longer true, and except for special angles of incidence, the reflection matrix depends on both Rashba and Dresselhaus strengths. The general reflection matrices are derived in the appendix. In this paper we will take the limit of $\pi/4$ reflections, that leads to simple, spin-orbit independent matrix elements.

The resulting beams follow path $I$ ($II$) that consists of a first horizontal ${\cal L}_I$ (vertical ${\cal L}_{II}$) arm made of Rashba-Dresselhaus medium whose length is $L_I$ ($L_{II}$). The electrons are then specularly reflected from an ideal mirror $M_1$ ($M_2$), that also mixes spin directions, followed by a vertical ${\cal L'}_I$ (horizontal ${\cal L'}_{II}$) arm of length $L_I$ ($L_{II}$) of the same material. The mirrors can be implemented as a simplified version of the beam splitters of reference \cite{OliverYamamoto}. Then the electrons pass through a second QPC (${\rm QPC_2}$) described by the corresponding S-Matrix $S_2$. Finally, two electron beams are collected at detector $D_i$ ($i=1,2$), and we have $\Psi_{D_i}= \Psi_{I,i}+\Psi_{II,i}$, where, $\Psi_{I,i}$ ($\Psi_{II,i}$) is the corresponding transferred spinor through the i{\it th}-arm. These amplitudes can be written in terms of the injected spinor $\Psi_0$ as $\Psi_{D_i}={\cal U}_{D_i}\Psi_0$, where the $2\times 2$ matrices ${\cal U}_{D_i}$ (generalized comparator operators \cite{Peskin}) are given by
\begin{eqnarray}\label{phase1}
\fl{\cal U}_{D_1}=(t_2) \exp{\Big [\frac{i}{\hbar}\int_{{\mathcal L'}_I}{d {\bf l}\cdot}({\bf p}-e{\bf A}-g\bfa^a\tau^a)\Big]} (r_l)\exp{\Big[\frac{i}{\hbar}\int_{{\mathcal L}_I}{d {\bf l}\cdot}({\bf p}-e{\bf A}-g\bfa^a\tau^a)\Big ]}(t_1)+\nonumber\\
(r_{2l}) \exp{\Big [\frac{i}{\hbar}\int_{{\mathcal L'}_{II}}{d {\bf l}\cdot}({\bf p}-e{\bf A}-g\bfa^a\tau^a)\Big]} (r_r)\exp{\Big[\frac{i}{\hbar}\int_{{\mathcal L}_{II}}{d {\bf l}\cdot}({\bf p}-e{\bf A}-g\bfa^a\tau^a)\Big ]}(r_{1l}),\nonumber\\
\fl{\cal U}_{D_2}=(r_{2r}) \exp{\Big [\frac{i}{\hbar}\int_{{\mathcal L'}_{I}}{d {\bf l}\cdot}({\bf p}-e{\bf A}-g\bfa^a\tau^a)\Big]} (r_l)\exp{\Big[\frac{i}{\hbar}\int_{{\mathcal L}_I}{d {\bf l}\cdot}({\bf p}-e{\bf A}-g\bfa^a\tau^a)\Big ]}(t_1)+\nonumber\\
(t_2) \exp{\Big [\frac{i}{\hbar}\int_{{\mathcal L'}_{II}}{d {\bf l}\cdot}({\bf p}-e{\bf A}-g\bfa^a\tau^a)\Big]} (r_r)\exp{\Big[\frac{i}{\hbar}\int_{{\mathcal L}_{II}}{d {\bf l}\cdot}({\bf p}-e{\bf A}-g\bfa^a\tau^a)\Big ]}(r_{1l}).\nonumber\\
\end{eqnarray}
Such operators applied to the initial state do not change the energy expectation value. The transmission and reflection matrices regarding both Rashba and Dresselhaus interactions, for $\pi/4$ incidence angle, are given by
\begin{equation}\label{reflectionmatrix}
(t_j)=\left (
 \begin{array}{cc}
    t_j  & 0\\
    0 & t_j
\end{array}\right )~~;~~
(r_{j[l,r]})=\frac{\sqrt{2}}{2}\left (
 \begin{array}{cc}
    r_j  & \pm ir_j\\
    \pm ir_j & r_j
\end{array}\right ),
\end{equation}
where the subscripts $j$ correspond to the beam splitter index (see figure \ref{fig1}) and $r,l$ (corresponding to $+,-$ in the non diagonal matrix elements, respectively)  encode whether the electron current is reflected counter-clockwise ($l$) or clockwise ($r$). $r_j$ and $t_j$ are the reflection and transmission coefficients for the $j-$th beam splitter, while for the mirrors, the reflection coefficients are equal to 1. Note that ${\cal U}_{D_i}$ is not a unitary operator. The normalization condition $|\Psi_{D_1}|^2+|\Psi_{D_2}|^2=1$ for the total probability at the detectors require that ${\cal U}^{\dagger}_{D_1}{\cal U}_{D_1}+{\cal U}^{\dagger}_{D_2}{\cal U}_{D_2}=\nbOne$, the unit matrix. This simply means that the amplitudes received at the detectors do not interfere. The arms of the interferometer can be built from gate defined quasi one dimensional paths implemented on a 2DEG, where all transport is kept within one of the available transverse modes. The scattering length is assumed to be long enough, so that phase relations can be accurately described by the path lengths and the spin-orbit strengths as in the Datta Das\cite{DattaDas} switch arrangement. 

\section{Results: Spin diagonal mirrors and beam splitters}

In this section we consider a simplified version of the filtering device where beam splitters and mirrors are considered diagonal matrices or scalars. Although this approximation does not contemplate the matrix nature of the reflections we will obtain a simple scenario for the filtering properties of the device. The full problem will be treated below where essentially the same qualitative results are obtained.

If the electric field ${\bf E}$ is uniform and static, the operators ${\bf p}-e{\bf A}$ and $g\bfa^a\tau^a$ commute. Thus, we can separate the {\it orbital} from the {\it internal} translation operators. For simplicity we will assume a square interferometer, thus $L_I=L_{II}=L$. Otherwise there are no restrictions or approximations related to the dimensions of the arms of the interferometer. As in Chen and Chang \cite{SHChen} we will make the discussion general by treating both the Rashba and Dresselhaus spin-orbit coupling on equal footing.

Concerning the {\it orbital} contribution, it is easy to see that this will consist of a global phase $\exp[{{\bf p} \cdot ({\bf L_1+L_2})}]$ which we can drop, and a relative $U(1)$ phase $\varphi_B$ which arises from the noncommutation of ${\bf p}$ and ${\bf A}$. Using the definition for the magnetic flux $\Phi_B=BL^2$ and that for the flux quantum $\phi_0=h/e$, the nontrivial {\it orbital} phase  is written as $2\pi\varphi_B=2\pi\Phi_B/\phi_0$. On the other hand, the internal part gives rise to the $SU(2)$ spin-dependent phase contribution. In order to simplify the resulting expressions, we introduce the adimensional variable
\begin{equation}\label{Lambda}
\Lambda=(m^*L/\hbar)\sqrt{\alpha^2+\beta^2}, 
\end{equation}
that will be the crucial control parameter governing the SO interaction. Furthermore, we introduce the definitions $\theta\equiv\tan^{-1}(\beta/\alpha)$ along with the matrices $\tilde{\sigma}^1\equiv\cos\theta\sigma^x-\sin\theta\sigma^y$ and $\tilde{\sigma}^2\equiv\sin\theta \sigma^x-\cos\theta \sigma^y$, such that $(\tilde{\sigma}^i)^2=\nbOne$, with $\nbOne$ the identity matrix in spin space. After the previous considerations we can rewrite equation \ref{phase1} in the form 
\begin{eqnarray*}
 {\cal U}_{D_1}&=&(t_2)\exp({-i\Lambda\tilde{\sigma}^1})(r_l)\exp({-i\Lambda\tilde{\sigma}^2})(t_1)+\nonumber\\ 
&& \exp({2\pi i\varphi_B})(r_{2r})\exp({-i\Lambda\tilde{\sigma}^2})(r_r)\exp({-i\Lambda\tilde{\sigma}^1})(r_{1l}),\\
 {\cal U}_{D_2}&=&(r_{2r}) \exp({-i\Lambda\tilde{\sigma}^1})(r_l)\exp({-i\Lambda\tilde{\sigma}^2})(t_1)+\nonumber\\
&& \exp({2i\pi\varphi_B})(t_2)\exp({-i\Lambda\tilde{\sigma}^2})(r_r)\exp({-i\Lambda\tilde{\sigma}^1})(r_{1l}).
\end{eqnarray*}
Due to the symmetry of these expressions (${\cal U}_{D_2}$ is obtained from ${\cal U}_{D_1}$ by the substitutions $r_2\leftrightarrow t_2$) we can focus on the first process, and obtain the second by making the necessary substitutions. Using the identity $\exp(\pm i \gamma \sigma^n)=\cos \gamma\nbOne \pm i\sigma^n\sin\gamma$, valid also for our redefined $\tilde{\sigma}$, the matrix ${\cal U}_{D_1}$ takes the form
\begin{eqnarray*}
{\cal U}_{D_1}&=&t_1 t_2 [\cos^2\Lambda\nbOne-i\sin\Lambda\cos\Lambda(\tilde{\sigma}^1+\tilde{\sigma}^2)-\tilde{\sigma}^1\tilde{\sigma}^2\sin^2\Lambda]+\nonumber\\ 
&& r_1 r_2 e^{{2i\pi\varphi_B}}[\cos^2\Lambda\nbOne-i\sin\Lambda\cos\Lambda(\tilde{\sigma}^1+\tilde{\sigma}^2)-\tilde{\sigma}^2\tilde{\sigma}^1\sin^2\Lambda].
\end{eqnarray*}
Now, we can easily determine that $\tilde{\sigma}^1\tilde{\sigma}^2=\sin 2\theta\nbOne-i{\sigma}^z\cos 2\theta$ thus $\tilde{\sigma}^2\tilde{\sigma}^1=\sin2\theta\nbOne+i{\sigma}^z\cos 2\theta$ and $\tilde{\sigma}^1+\tilde{\sigma}^2=(\cos\theta+\sin\theta)(\sigma^x-\sigma^y)$. Substituting these results and rearranging the obtained expressions leads to 
\begin{displaymath}
{\cal U}_{D_1}={\mathcal A}_{+}[\cos^2\Lambda-\sin^2\Lambda\sin 2\theta]\nbOne+i\sin\Lambda \nbM, 
\end{displaymath}
where we have introduced the traceless matrix $\nbM={\mathcal A}_{-}\sin\Lambda\cos 2\theta\sigma^z-{\mathcal A}_{+}\cos\Lambda(\cos\theta+\sin\theta)(\sigma^x-\sigma^y)$ and ${\mathcal A}_{\pm}=t_1t_2\pm r_1r_2 e^{2i\pi\varphi_B}$. The traceless condition simplifies the diagonalization of $\nbM$, and the eigenvalues for ${\cal U}_{D_1}$ are easily found to be
\begin{eqnarray}\label{evalue1}
&&\fl \lambda^{D_1}_{\pm}={\mathcal A}_{+}[\cos^2\Lambda-\sin^2\Lambda\sin 2\theta]\mp i\sin\Lambda%\times\nonumber\\
\sqrt{{\mathcal A}^2_{-}\sin^2\Lambda\cos^2 2\theta+2{\mathcal A}^2_{+}\cos^2\Lambda(1+\sin 2\theta)}.
 \end{eqnarray}
If we now define ${\mathcal B}_{\pm}=t_1 r_2\pm r_1 t_2 e^{2i\pi\varphi_B}$, the eigenvalues of the matrix ${\cal U}_{D_2}$ are obtained from the previous result by making the substitution ${\mathcal A}_{\pm}\rightarrow {\mathcal B}_{\pm}$
\begin{eqnarray}\label{evalue2}
&&\fl \lambda^{D_2}_{\pm}={\mathcal B}_{+}[\cos^2\Lambda-\sin^2\Lambda\sin 2\theta]\mp i\sin\Lambda%\times\nonumber\\
\sqrt{{\mathcal B}^2_{-}\sin^2\Lambda\cos^2 2\theta+2{\mathcal B}^2_{+}\cos^2\Lambda(1+\sin 2\theta)}.
 \end{eqnarray}
In order to get more insight into the nature of the conditions for perfect spin filtering we will specialize the previous expression to symmetric beam splitters i.e. $r_1=r_2=r$, and $t_1=t_2=t$. Within this case, we have $\mathcal{A}_{\pm}=t^2 \pm r^2e^{2i\pi\varphi_B}$. Since we are interested in filtering one spin component, say the up component, we now proceed to determine the vanishing conditions of the corresponding eigenvalue $\lambda^{D_1}_{+}$. 

From expressions (\ref{evalue1}, \ref{evalue2}), these vanishing conditions can be found by either having $\cos\Lambda=0$ or $\cos\Lambda\ne0$ (see also equation \ref{Lambda}). Although the former condition is mathematically only a particular case of the general solution, we distinguish it because the corresponding ${\cal U}_{D_1}$ becomes diagonal with respect to the original quantization axis, so we can speak of filtering along a {\it non-tilted} axis. Such a solution is also the simplest from the detection point of view since it involves the choice of a single quantization axis for the whole setup. The second condition ($\cos\Lambda\ne0$) corresponds to finding a new axis where the up spin is filtered and we  call such axis the {\it tilted} quantization axis. Note that both these filtering conditions (non-tilted and tilted) are {\it independent of the polarization axis and the energy of the incoming state}. We will comment further on this below.

\subsection{Non-tilted filtering}
\begin{figure}
\begin{center}
\includegraphics[width=7 cm]{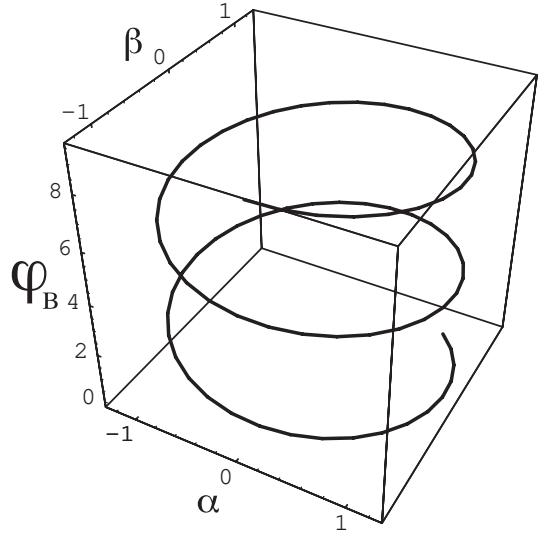}
\end{center}
\caption{Perfect filtering for the non-tilted axis (original incoming basis). The plot shows the relation between $\alpha$, $\beta$ in units of $\hbar/(m^* L)$, and $\varphi_B$ that yields perfect polarization of the spin from an unpolarized input. The figure corresponds the values $n,l=0$ according to equation \ref{nontilted1}.}
\label{fig2}
\end{figure}

Let us first analyze the {\it non-tilted} situation. In this case the filtering condition requiring $\lambda_+^{D_1}=0$ for all incoming energies (see equation \ref{evalue1}), leads to the relation
\begin{displaymath}
\tan 2\theta=-\frac{i(t^2 - r^2 e^{2i\pi\varphi_B})}{(t^2 + r^2 e^{2i\pi\varphi_B})}. 
\end{displaymath}  
Two $50-50$ beam splitters for which $r=i/\sqrt 2$, $t=1/\sqrt 2$, will then lead to the relation $\sin{\pi\varphi_B}\sin 2\theta=\cos{\pi\varphi_B}\cos 2\theta$, equivalent to the simple expression 
$\cos({\pi\varphi_B+2\theta})=0$, satisfied by the condition
\begin{equation}\label{nontilted}
\pi\varphi_B+2\theta=(2n+1)\frac{\pi}{2},
\end{equation}
where $n$ is an integer. Figure \ref{fig2} depicts the relation between the spin-orbit parameters and the magnetic flux, for $n,l=0$, necessary for perfect filtering of the up component in the original quantization axis. The spin-orbit parameters are in a reasonable range, as depicted in the figure, since for a GaAs heterostructure $\hbar\alpha\sim 3.9\times 10^{-12}{\rm eV~ m}$\cite{DattaDas}, $\hbar\beta\sim 2.4\times 10^{-12} {\rm eV~m}$ and $\hbar^2/m^* L\sim 1.7 \times 10^{-12}{\rm eV~ m}$, assuming the arm of the interferometer $\sim 1 \mu m$ and an effective mass of $m^*=0.046 m_0$. These parameters yield $|\alpha|, |\beta| < 6$ in units of $\hbar/(m^* L)$. Note that our definition of $\alpha,\beta$ differs by a factor $\hbar$ to the standard definition (see equation \ref{H1}). In reference\cite{MillerGoldhaberGordon} it is shown that gate control can vary $\alpha$ and $\beta$ parameters by a factor of 6 by applying gate voltages in the hundreds of mV.

The solutions are on a helix, as can be shown from the previous relations where
\begin{eqnarray}\label{nontilted1}
\alpha&=&\frac{\hbar}{m^* L}\sqrt{(2l+1)\pi/2}\cos[\pi/4(2n+1-2\varphi_B)],\nonumber\\ 
\beta&=&\frac{\hbar}{m^* L}\sqrt{(2l+1)\pi/2}\sin[\pi/4(2n+1-2\varphi_B)].
\end{eqnarray}
The integer $n$ was defined in equation \ref{nontilted} while the second integer $l$ is defined by the condition $\cos{\Lambda}=0$.

\begin{figure}
\begin{center}
\includegraphics[width=9 cm]{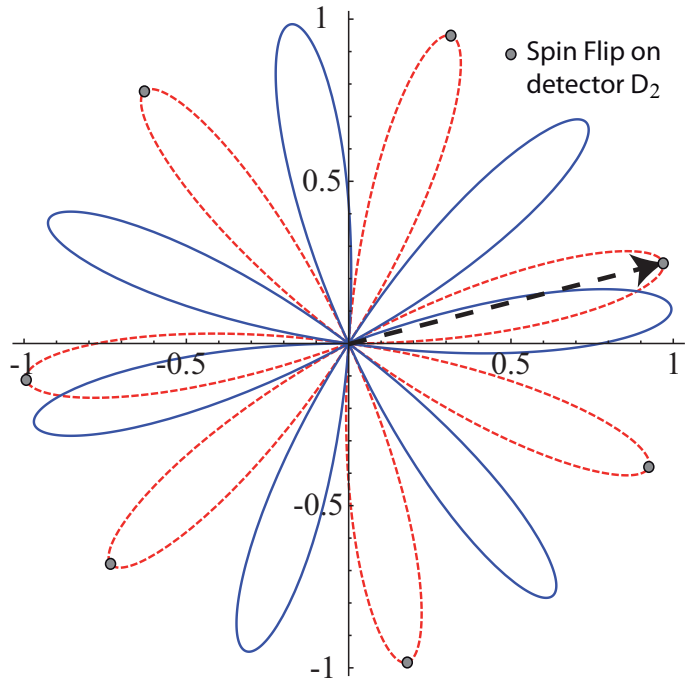}
\end{center}
\caption{Filtering probability for the non-tilted solution of detector $D_1$ for $n=0$,  $l=0$ solid (blue) curve and n=1, l=0 dashed (red) line. The radius vector depicted shows the filtered probability for the output spinor for one whole period in the parameters $\alpha,\beta$ as given in the figure \ref{fig2}. The position of the dashed vector corresponds to $\varphi_B=0.25$. The grey points represent ``spin flipping" or opposite filtering solutions for detector $D_2$.}
\label{fig3}
\end{figure}

The previous conditions, depicted in figure \ref{fig2}, do not tell us about the intensity of the signal received in detector $D_1$ i.e. the efficiency of the filter given an incident intensity. For this, one has to look back at the eigenvalues. While $\lambda^{D_1}_+=0$ the amplitude of the outgoing polarized spinor at detector $D_1$ is given by
\begin{equation}\label{outputnontilted}
\Psi_{D_1}={0 \choose \lambda^{D_1}_-\psi^-_0}={0 \choose i e^{i\pi\varphi_B}\cos({\pi\varphi_B-2\theta})\psi^-_0},
\end{equation}
whose modulus squared is $\cos^2(\pi\varphi_B-2\theta)|\psi^-_0|^2$. Figure \ref{fig3} shows a polar plot for the amplitude of the filtered signal (radius vector) as a function of the parameter designating the field flux $\varphi_B$ and the $\alpha,\beta$ combination given by equation \ref{nontilted1} for $n=0,1$ and $l=0$. The figure shows that while filtering occurs for all the fluxes (given the appropriate values of $\alpha,\beta$) the amplitude can be zero, or very small, for some flux values i.e. in this case, the detector $D_2$ gets most of the total amplitude. On the other hand, for some values of the flux, filtering can be very strong since the probability for a polarized spin can approach unity.
The behavior of the second detector $D_2$, while the first detector sees a filtered signal,  can be obtained through the eigenvalues of that detector having substituted the condition $\lambda_+^{D_1}=0$, namely
\begin{eqnarray} 
&&\lambda^{D_2}_{+}=-ie^{i\pi\varphi_B},\nonumber\\
&&\lambda^{D_2}_{-}=ie^{i\pi\varphi_B}\sin({\pi\varphi_B-2\theta}). 
\end{eqnarray}
It is obvious that the second detector $D_2$ does not filter concomitantly with the $D_1$ in general. Furthermore, one can only find conditions for the second component to be zero (opposite filtering to detector $D_1$) since the first component has modulus one. This takes us to the non-tilting {\it spin flipped} or opposite filtering solution at detector $D_2$, only occurring while detector $D_1$ is filtering with maximal efficiency i.e. maximal polar radii in figure \ref{fig3}. 

The filtering amplitude is proportional to the projection of the incoming spinor (which has arbitrary weights onto the chosen quantization axis) to the surviving component at the output (see equation \ref{outputnontilted}). This means that for each arbitrary incident spinor from the Fermi sea one gets a filtering probability that depends on this projection. The resulting polarized current will thus have a random noise associated with this effect besides the contribution from shot noise. 

It is important to note that this solution does not appear in Abelian approximation (only exact in the case $\alpha^2=\beta^2$ and in one dimension) to the translation operator, where the $SU(2)$ gauge vector operator has the same algebra as the $U(1)$ gauge vector. The previous approximation was implemented in reference \cite{SHChen} by neglecting the commutator between components of the $SU(2)$ gauge vector within a finite difference scheme. In this sense, the non-tilted case is an intrinsically non-Abelian scenario for spin filtering.

\subsection{Tilted filtering axis}\label{idealtilted}
The tilted axis filtering scenario was discussed, within the tight-binding model, by Hatano, Shirasaki and Nakamura\cite{Hatano} when the Rashba coupling is present. In their approach, the interferometer involves an incoming lead and one outgoing lead, in contrast to our Mach-Zehnder configuration. The non-Abelian treatment is exact within their model, and requires a tilted outgoing axis to realize perfect spin filtering.

For the Mach-Zehnder configuration, addressed here, the {\it tilted} axis solution (i.e. $\cos\Lambda \neq 0$), requires $\lambda^{D_1}_+=0$, which implies
\begin{eqnarray*}
&&~~~\fl {\mathcal A}_{+}[\cos^2\Lambda-\sin^2\Lambda\sin 2\theta]= i\sin\Lambda\nonumber
\sqrt{{\mathcal A}^2_{-}\sin^2\Lambda\cos^2 2\theta+2{\mathcal A}^2_{+}\cos^2\Lambda(1+\sin 2\theta)}. 
\end{eqnarray*}
Squaring both sides and after some algebra one finds 
\begin{equation}\label{casi1}
{\mathcal A}^2_{+}=\sin^4\Lambda\cos^2 2\theta({\mathcal A}^2_{+}-{\mathcal A}^2_{-}).
\end{equation}
Using the definitions for ${\mathcal A}_{\pm}$, and taking the square root, we reduce equation \ref{casi1} to
\begin{displaymath}
t^2+r^2 e^{2i\pi\varphi_B}=2rte^{i\pi\varphi_B}\sin^2\Lambda\cos 2\theta.
\end{displaymath}
Employing the $50-50$ mirror condition, we get after substitution
\begin{equation}
\label{nontilted5050}
\sin\pi\varphi_B=\sin^2\Lambda\cos 2\theta.
\end{equation}
\begin{figure}
\begin{center}
\includegraphics[width=15 cm]{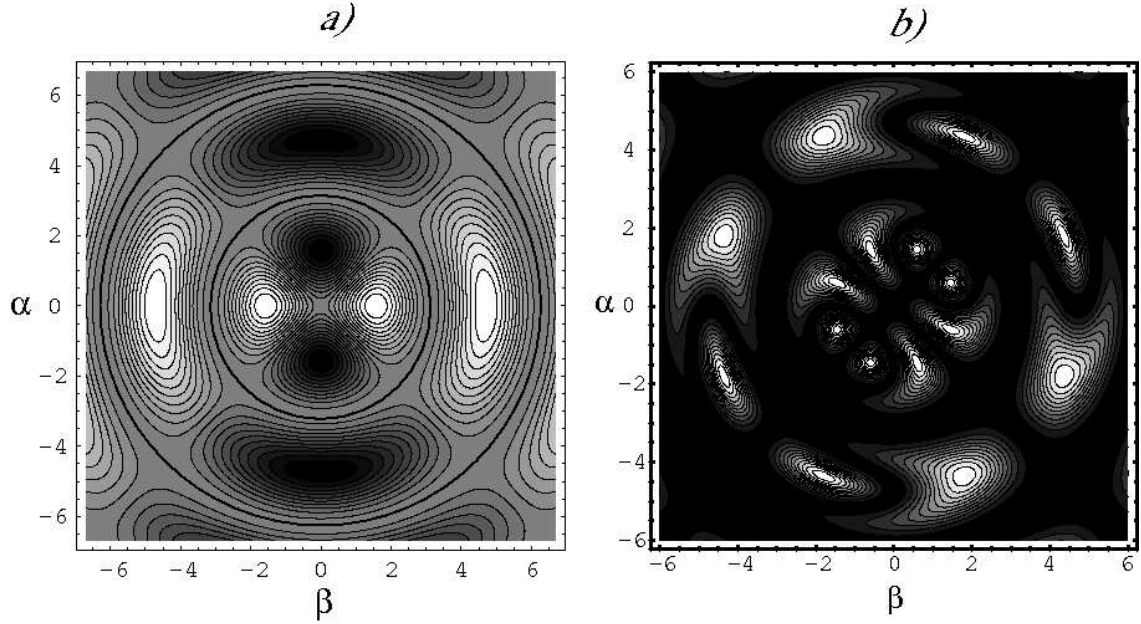}
\end{center}
\caption{ a) Perfect filtering by interference for the tilted axis. The plot shows the relation between $\alpha$, $\beta$ in units of $\hbar/(m^* L)$, and $\sin\pi\varphi_B$ in a contourplot, the darker regions indicate larger values for the magnetic flux needed to yield perfect filtering, from an unpolarized input. Highlighted circles depict the zero flux solutions that yield perfect filtering. b) Perfect filtering probability for the tilted axis. The plot shows the relation between $\alpha$, $\beta$ in units of $\hbar/(m^* L)$, and the filtered intensity in a contourplot. The lighter regions indicate larger values for the intensity of filtering for the relation between parameters depicted in figure \ref{fig45}a. Note that the circles evident from figure \ref{fig45}a correspond to zero output amplitude.}
\label{fig45}
\end{figure}
%\begin{figure}
%\begin{center}
%\includegraphics[width=9cm]{Figure5.eps}
%\end{center}
%\caption{Perfect filtering probability for the tilted axis. The plot shows the relation between $\alpha$, $\beta$ in units of $\hbar/(m^* L)$, and the filtered intensity in a contourplot. The lighter regions indicate larger values for the intensity of filtering for the relation between parameters depicted in figure \ref{fig4}a. Note that circles and diagonals evident from figure \ref{fig45}a correspond to zero output amplitude.}
%\label{fig5}
%\end{figure}

This is the relation between the spin-orbit parameters and the magnetic flux that leads to perfect filtering in the tilted axis. The solution is depicted in a contourplot in figure \ref{fig45}a where the value of $\sin\pi\varphi_B$ is represented in shades of gray as a function of $\alpha$ and $\beta$. Each contour corresponds to a constant magnetic flux value and runs over the perfect filtering values of $\alpha$ and $\beta$. The circular contour, depicted in the figure, corresponds to a $\varphi_B=0$ solution to equation \ref{nontilted5050} that leads to $(m^*L/\hbar)\sqrt{\alpha^2+\beta^2}=p\pi$, for $p$ integer. The figure depicts the solution for $p=1,2$, i.e. circles in units of $\hbar/(m^*L)$.  

In order to see if the filter is actually working, we must address the filtered amplitudes by looking to the second eigenvalue at detector $D_1$. For the filtering condition
\begin{equation}
\lambda^{D_1}_{-}=-2ie^{i\pi \varphi_B}\sin\pi \varphi_B\left[ \cos^2\Lambda-\sin^2\Lambda\sin 2\theta\right].
 \end{equation}
Substituting equation \ref{nontilted5050} in this expression and computing the modulus squared of the eigenvalue, we determine the strength of the filtered output, as was done in equation \ref{outputnontilted}. We have depicted the analytical solution for a range of values of $\alpha,\beta$ in the contour plot of figure \ref{fig45}b. The darkest shade corresponds to zero amplitude, and as the shade lightens the probability is higher for the filtered output. We note that the filtering solutions for the circular contours in figure \ref{fig45}a and the lines $\alpha=\pm\beta$ have zero amplitude. Such zero amplitude solutions correspond to those of  ``localized solutions'' of  Cheng and Chang\cite{SHChen} where there is no filtered output. 
\begin{figure}
\begin{center}
\includegraphics[width=8 cm]{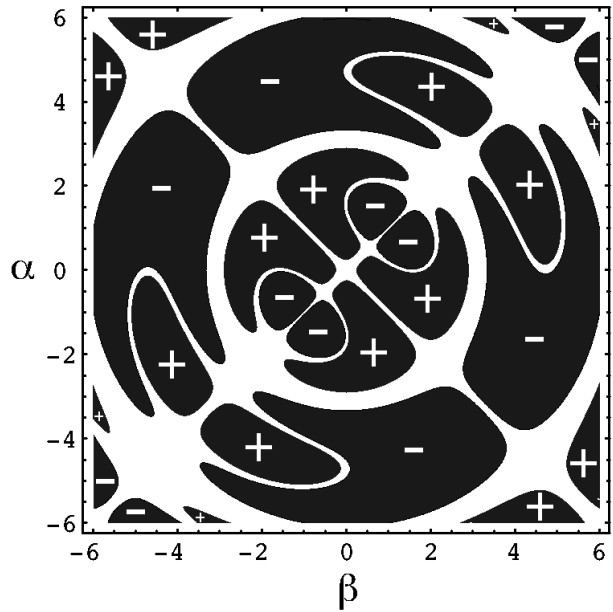}
\end{center}
\caption{Detector $D_2$ output while $D_1$ filters out the up spin component (spin down polarization). The plus (minus) zones represent the regions where only the up spin (down) survives at the $D_2$ detector. Note that either one or the other is filtered. The white regions represent no output in the detector and correspond to the localized phase. On can have either up or down spin filtering in $D_2$ while up spin is filtered out in $D_1$.}
\label{fig6}
\end{figure}
Behavior of detector $D_2$, while $D_1$ is filtering out the spin up component (spin down polarization), is shown in figure \ref{fig6}. Regions with plus (minus) signs depict up (down) spin phases for detector $D_2$. Note that the two regions are mutually exclusive so that while pure spin down is being detected in $D_1$ one can have either spin up or spin down in $D_2$ depending on the range of $\alpha,\beta$. The white regions correspond to no output at $D_2$. Comparing with figure \ref{fig45}b we see that no-output region are not identical for both detectors, these being larger for $D_1$, i.e. one can have zero output at $D_1$ while having non-zero output at $D_2$. As discussed before, the outputs depicted in figure \ref{fig6} are also modulated by the magnitude of the corresponding component at the input, so the probability of the output exhibits noise coming from the random input spin orientation.
\begin{figure}
\begin{center}
\includegraphics[width=12cm]{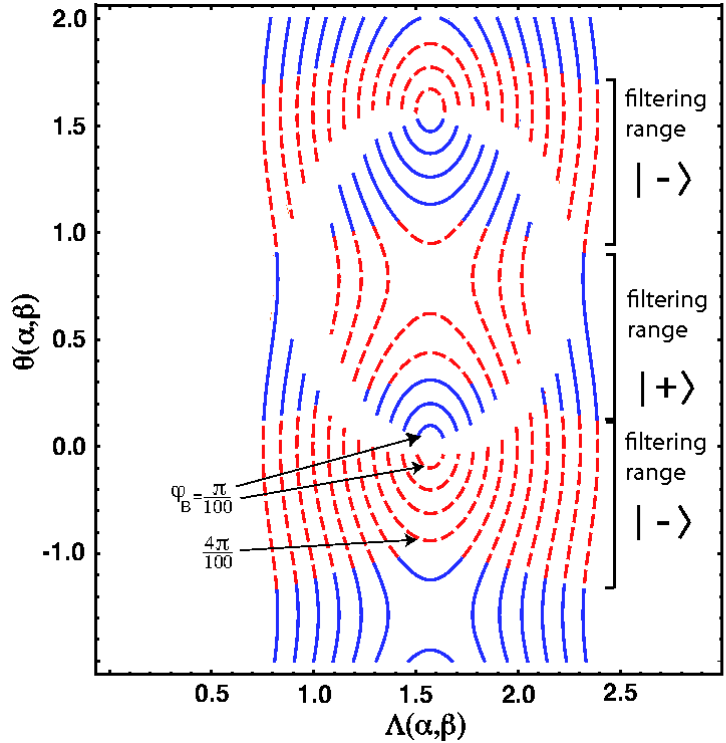}
\end{center}
\caption{The zeroes of the first (dashed line or red online) and second (solid line, blue online) eigenvalues of ${\cal U}_{D_2}$. When the first eigenvalue vanishes  (and the second is non-zero), for specific combinations of $\alpha,\beta ~{\rm and}~\varphi_B$  the interferometer produces a perfectly polarized output in the $|- \rangle$ state. Only a particular discrete set of solutions for $\varphi_B$ is depicted.}
\label{fig7}
\end{figure}

\section{Non diagonal mirror and beam splitter reflections}
Including the non diagonal matrix character of reflections at mirrors and beam splitters shifts the operation parameters of the spin filter but yields essentially the same qualitative results. The conditions must now be derived numerically. We start from equation \ref{phase1} with the transmission and reflection matrices in equation \ref{reflectionmatrix}. For the particular choice of $\pi/4$ incidence on the mirrors (see Appendix), the particularly simple non-tilting scenario described above is not possible. The extra parameter given by the angle of incidence on the mirrors/beam splitters lends itself to making this regime accessible, but we will not pursue it here. The more general scenario of a tilted axis yields a whole range of possible filtering solutions. 

Diagonalizing ${\cal U}_{D_2}$ in equation \ref{phase1} we find two eigenvalues. Setting the first eigenvalue to zero implies that in this rotated space the spinor is fully polarized (one of the entries of the output spinor is zero) as described in equation \ref{outputnontilted}. Setting this eigenvalue to zero means setting its real and imaginary parts to zero. Such zeroes are depicted in figure \ref{fig7} by the dashed lines (red online) for different values of the magnetic field and specific combinations $\Lambda(\alpha,\beta)$, defined in equation \ref{Lambda}, and $\theta=\tan^{-1}(\beta/\alpha)$. In order for filtering to be performed such zeroes must be accompanied by non-zero values of the second eigenvalue in the same detector. The zeroes of the second eigenvalue are depicted in figure \ref{fig7} by the solid lines (blue online) which are non-overlapping with the dashed lines for the first eigenvalue. Thus the figure shows alternative filtering conditions for either spin up or spin down in the tilted basis. 

The circular empty region in the middle of the plot correspond to non-polarized output in the tilted axis. Such a region contains some pointlike solutions that are of less interest experimentally since they would be difficult to tune. We recall that the previous discussion in section \ref{idealtilted} is equally valid in this case, all incoming electrons at the input are polarized at the output no matter their energy as long as particular parameters ranges in the $\alpha,\beta,\Phi_B$ space are met. 
\begin{figure}
\begin{center}
\includegraphics[width=15cm]{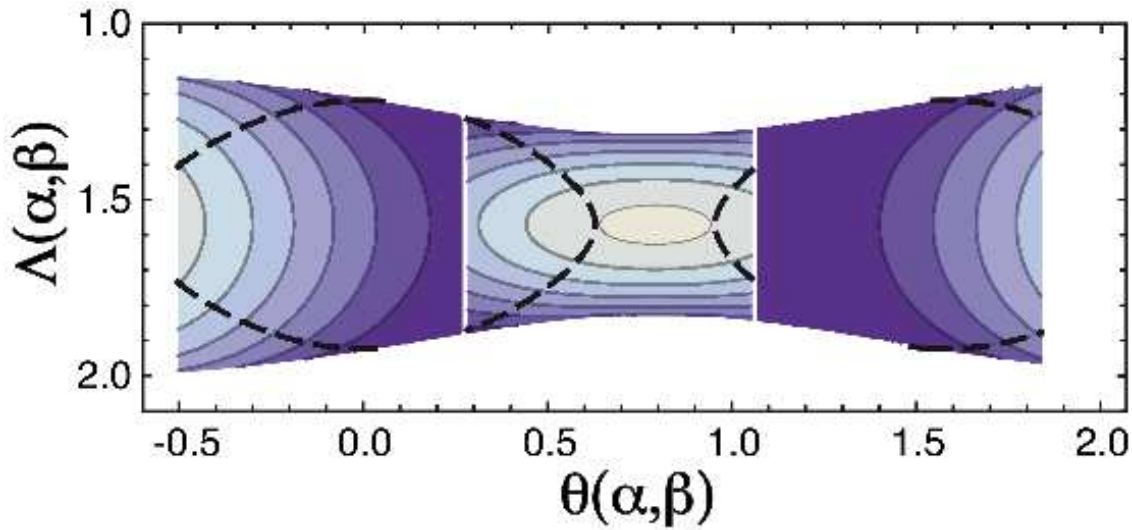}
\end{center}
\caption{The dashed curves represent zeroes of the first eigenvalue for $\varphi_B=5 \pi/100$ upon a contourplot for the modulus of the second eigenvalue. The lighter shades represent higher values of the output polarization. One can extract the SO strengths from the plot by solving a simple system of equations for each value read off on the dashed curves.}
\label{fig8}
\end{figure}

In order to see the magnitude of the spin polarization for a particular value of the external magnetic field we draw a contour map of the magnitude of the second eigenvalue while the first one is zero. The background value at the dashed curves in figure \ref{fig8}, show the intensity of the pure down spin polarization at detector $D_2$ when at $\varphi_B=5\pi/100$. The highest values of output achieved corresponds to the lighter shades on the contourmap.  

\section{Summary}
We have proposed a perfect spin filtering device based on a Mach-Zehnder type spin interferometer. The regimes of operation are subject to no limitations on the spin-orbit strengths and interferometer dimensions as in previous work. The treatment can be easily extended to unequal arm lengths and angles of incidence on the mirrors/beam splitters, that are likely to occur in the actual implementation of the interferometer. Such a generalization would provide additional parameters to manipulate filtering conditions. In the simpler analysis above involving scalar mirrors, we find both a non-tilted and tilted axis spin filtering solutions referred to the axis of quantization in which one writes the input states and for arbitrary incoming energies. The non-tilted case is not found in the scenario where the $SU(2)$ gauge field is approximated by a $U(1)$ like gauge, and is peculiar to the full non-Abelian treatment. This solution has the advantage of simplicity. On the other hand, the tilted axis solutions are shown to be well approximated by the Abelianized forms of reference \cite{SHChen} valid for certain reasonable conditions of SO strengths in relation to the interferometer arm lengths. When realistic mirrors/beam splitters are introduced, the mixing of the spinor components leads only to non-tilted solutions when $\pi/4$ reflections are contemplated. In this situation we run out of adjustable parameters to tune a non-tilted solution, that should be recovered when other incidence angles are considered. The qualitative scenarios for the operation of the diagonal and  non-diagonal mirrors are the same and only the parameter combinations for filtering change.

Perfect filtering means that all spins in one of the detectors are polarized always in the same axis and orientation. This has the drawback that the current is not steady since the probability of producing a completely polarized electron varies with the initial projection, of the input spinor, onto the chosen quantization axis. This projection is random as electrons are injected from the Fermi sea\cite{Yamamoto}. A density matrix approach should be implemented so that one can also assess finite temperature effects on the filter operation. It should be also noted that the interference setup does not produce a pure spin current, since polarization is accompanied by a charge current.

An interesting insight, exploiting the analogy with the Aharonov-Bohm effect in the Abelian case, comes from observing the role of $\Lambda$ in the non-Abelian case. $\Lambda$ and the voltage $V$ essentially play the same role as the pair $2\pi\varphi_B$ and magnetic flux. Indeed, for a purely Pauli type SO interaction, as $\Lambda=(mL/{\hbar})\alpha$ and $\alpha=\hbar eE/(m^2 c^2)$, then $\Lambda$ can be rewritten as $2\pi E L/(2\pi m c^2/e)=2\pi V_E/V_0$, where $V_E=EL$, the voltage along the arm of length $L$ in an electric field of strength $E$. $V_0$ is a quantum of voltage\cite{Medina}. Although $V_0$ is very large for this calculation, the material Rashba coefficient would lower it to the order of $1~ eV/e$.

\ack
We acknowledge fruitful discussions with C. Chatelain, J. C. Egues and R. Winkler. This work was supported by CNRS-Fonacit grant PI-2008000272.

\section*{Appendix}

Here we derive the general conditions for reflection at a beam splitter on a mirror in the presence of both Rashba and Dresselhaus interactions. Starting from Hamiltonian in equation \ref{Hamiltonian} we can solve exactly for the eigenvalues and eigenfunctions. Ignoring the Zeeman term we have
\begin{figure}
\begin{center}
\includegraphics[width=10cm]{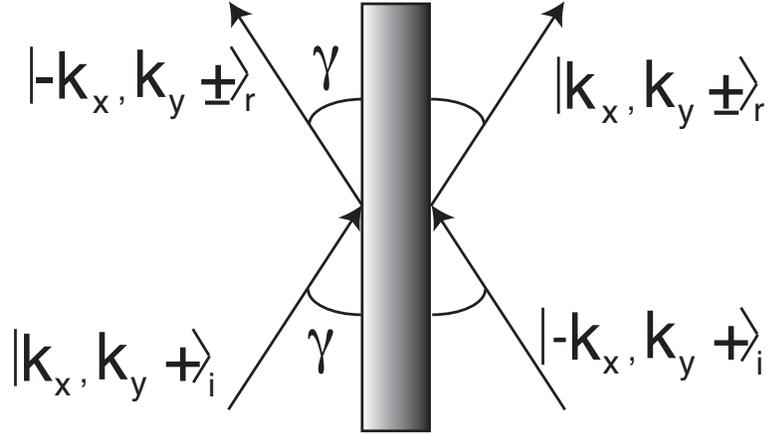}
\end{center}
\caption{Detector $D_2$ output while $D_1$ filters out the up spin component (spin down polarization). }
\label{figappendix}
\end{figure}

\begin{equation}
\varepsilon_{\pm}=\frac{\hbar^2k^2}{2m^*}\pm\sqrt{k^2(\alpha^2+\beta^2)+4 \alpha\beta k_x k_y},
\end{equation}
with eigenfunctions given by
\begin{equation}
\fl | {\bf k} ~\pm \rangle_{i}=\frac{1}{\sqrt{2}}\left (
 \begin{array}{cc}
    1\\
    \mp F(k_x,k_y)
\end{array}\right )~~,~~ F(k_x,k_y)=\frac{k_x(\beta-i\alpha)+k_y(\alpha-i\beta)}{\sqrt{k^2(\alpha^2+\beta^2)+4\alpha\beta k_x k_y}},
\end{equation}
where ${\bf k}=(k_x,k_y)$, $\pm$ stand for the two eigenvalues and the subindex $i$ stands for incident wave. The convention we take, according to the figure, is that $k_x~\rm{and} ~k_y$ are positive components for the incident electron. Referred to those components, one can obtain the reflected basis components by changing $k_x\rightarrow - k_x$ and $k_y\rightarrow k_y$ as the momentum in the $y$ direction is conserved. To obtain the projections in terms of the reflected basis we write
\begin{equation}
 | {\bf k} ~\pm \rangle_{i}= a_{\pm} | {\bf k} ~+ \rangle_{r}+b_{\pm}  | {\bf k} ~- \rangle_{r},
 \end{equation}
 where the subindex on the right indicates the reflected complete basis set. One can then compute the superposition coefficients $a_{\pm}$ and $b_{\pm}$ by performing the appropriate overlaps between incoming and outgoing wavefunctions
 \begin{eqnarray}
 a_{\pm} &=&  _r\langle k + |k \pm\rangle_i=1/2\left [1 \pm F^*(-k_x,k_y)F(k_x, k_y)\right ],\cr
 b_{\pm}&=&  _r\langle k - |k \pm\rangle_i=1/2\left [1 \mp F^*(-k_x,k_y)F(k_x, k_y)\right ].
 \end{eqnarray} 
 Each of the outgoing amplitudes gets multiplied by the scalar reflection coefficient $r$ in the case of the beam splitter and $r=1$ for perfect mirrors. The previous coefficients govern the ${\rm QPC_1}$, the upper reflection of ${\rm QPC_2}$ and $M_1$ in figure \ref{fig1}, while exchanges of $k_x\rightarrow -k_x$  would generate the corresponding matrix for the $M_2$ and the bottom reflection of ${\rm QPC_2}$. 
 
 The wavector components can be expressed as ${\bf k}=(k \sin\gamma,k \cos\gamma)$ for a generic incident angle as seen in the figure. For the case of $\gamma=\pi/4$, the reflection matrices are particularly simple and one obtains equation \ref{reflectionmatrix}, where the transmission matrix is trivially diagonal since the electron beam does not change direction.
 
 A coordinate independent way to state the general result is by identifying $F(k_x, k_y)=\exp{i\phi_i}$ and $F(-k_x, k_y)=\exp{i\phi_r}$ then one can write the full reflection/transmission matrix as
 \begin{equation}
 \fl \left (
 \begin{array}{cccc}
    r\cos[(\phi_r-\phi_i)/2]  & i r\sin[(\phi_r-\phi_i)/2] & t & 0\\
   i r\sin[(\phi_r-\phi_i)/2]  & r\cos[(\phi_r-\phi_i)/2]  & 0 & t\\
    t & 0 & r\cos[(\phi_r-\phi_i)/2] & -i r \sin[(\phi_r-\phi_i)/2]\\
    0 & t & -i r\sin[(\phi_r-\phi_i)/2] & r\cos[(\phi_r-\phi_i)/2]
\end{array}\right )
\end{equation}

\end{document}